\documentstyle[12pt]{article}
\textheight 
 235mm
\textwidth 
 155mm
\newcommand{\bea}{\begin{eqnarray}}
\newcommand{\eea}{\end{eqnarray}}
\newcommand{\ba}{\begin{array}}
\newcommand{\ea}{\end{array}}
\newcommand{\hh}{\hspace*{10mm}}
\newcommand{\hhh}{\hspace*{15mm}}
\newcommand{\h}{\hspace*{.5mm}}
\newcommand{\w}{\vspace*{3mm}}
\newcommand{\nn}{\nonumber}

\newcommand{\half}{ \frac{1}{2}}

\begin{document}

\begin{titlepage}
\newpage
\setcounter{page}{0}
\null

\rightline {hep-th/9605160}

\rightline{ZU-TH-15/96}

\rightline{May 1996}
\vspace{25mm}

\begin{center}
{\Large {\bf
  On the Holomorphic Gauge Quantization of 

\vspace{0.5cm}
the Chern-Simons  Theory and

\vspace{0,5cm}
 Laughlin Wave Functions}}

\vspace{1cm}

{\bf\large M.Eliashvili }
\vspace{0.5cm}

{\it Department of Theoretical Physics \\
Tbilisi Mathematical Institute\\
Tbilisi 380093 Georgia\\
E-mail address: simi@imath.acnet.ge}

\end{center}

\vspace{1cm}

\centerline{\bf Abstract}

\indent

 Chern-Simons-Matter Lagrangian with noncompact
gauge symmetry group is considered.  The theory  is quantized in the
holomorphic gauge with a complex gauge fixing condition.
The model is discussed, in which  the the gauge and matter fields
are accompanied by the complex conjugate counterparts. 
It is argued, that such a theory represents an adequate framework for the
description of the quantum Hall states.

\vspace{1cm}

\vspace*{\fill}
\end{titlepage}

\pagebreak

{\section {Introduction}}

One of the characteristic features of two-dimensional systems
is that the Green functions  (correlators) can be factorized
into  the product of holomorphic and antiholomorphic parts 
and corresponding gauge connections can take a complex values.

A known  example of a non-real gauge potential is provided by the integrable
connection over the configuration space arising from the Yang-Baxter equations.
This connection  is represented by the one-form \cite{Kohno}
\bea
\omega =const\cdot
 \sum_{I\not= J}T^a_{I}\otimes T^a_{J}  d\h \ln (z_I-z_J) \label{eq:Ko}
\eea
 where  $z_I =x_I+iy_I$ are the complex coordinates  and
 $T^a_{I}$ are the generators of symmetry group for the  $I^{th}$ particle.
This connection  governs the monodromy behavior of conformal
blocks  in (1+1) dimensional current algebra and enters into the Knizhnik - Zamolodchikov (KZ)
\cite{Knizhnik} equation
\bea
\Bigl(\frac{\partial}{\partial z_I}-\frac{1}{k+c}
 \sum_{I\not= J}\frac {T^a_{I}\otimes T^a_{J}}{ z_I-z_J }\Bigr)
\Psi (z_1,...,z_N)=0 \label{eq:KZ}
\eea
\bea
\frac{\partial}{\partial \bar z_I}\Psi (z_1,...,z_N)=0 \nn
\eea

The KZ connection plays an essential role in the  physics of
particles obeying the braid statistics  and in
the theory of quantum Hall effect (see {\it e.g.}\cite{Stone}). 
 In the later case the holomorphic
part of Laughlin wave function satisfies (\ref{eq:KZ}) 
and could be expressed as $N$-point correlation function
in certain conformal field theory 
 \cite {Moore}.

 The gauge potential (\ref{eq:Ko}) can be incorporated
into the framework of Chern-Simons (CS) gauge theory in 2+1 dimensions.
Formally, the problem reduces to the quantization  of the theory describing
the matter interacting with the C-S fields in the holomorphic
 gauge, where corresponding gauge condition is expressed by the
complex matrix equation 
 \bea
{\cal A}_x+i{\cal A}_y=0, \label{eq:GC}
\eea
 ( ${\cal A}_\mu$   is the Lie-algebra valued gauge connection)

Remind, that this type of gauge has been presented as a solution of a Gau{\ss} law
constraint in a discussions  of quantum holonomies \cite{Guadagnini},
and BRST quantization of non-abelian CS gauge theories \cite{Lee}.

Note, that holomorphic gauge quantization as considered in {\it e.g.}
 \cite{Lee} leads to non-Hermitean Hamilton operator and for consistency one has to introduce in the Hilbert   space a compensating integration measure,
   respectively which Hamiltonian  is self-adjoint \cite{Verlinde}.

It must be emphasized, that
 as well as the complex gauge condition is imposed in the CS theory with a compact gauge group and real gauge fields, equation (\ref{eq:GC})
must be understood in the sence of some analytic continuation.

It is worth pointing out at this point that 
in the paper \cite{Witten}  Witten had considered the theory with non-compact
(complex) gauge transformation group and complex  CS gauge fields.
 
  It was shown in this paper, that quantization os self-interacting
CS gauge fields can be performed as precisely as for compact groups, 
using standard tools and without any specific difficulties
(see also \cite{Bos,Elitzur}).

In the present note we consider the same scheme  as in \cite{Witten},
enlarging the system by the matter fields.
The  point of departure is the observation, that
in the holomorphic gauge   in
order to have real Lagrangian ({\it i.e.} unitary theory),
the matter fields as well  as the gauge degrees of freedom 
 must be accompanied by their complex conjugate
counterparts.
In the quantization procedure we follow Dirac's  classical method
\cite{Dirac}.

As the physical application we  
will try to give some convincing arguments, that the models  with
a complex gauge groups  can provide the    consistent description of
 the variety of  QHE wave functions.

\vspace{5mm}
\paragraph{ Outline and note on the conventions}

The paper is organized in following way.
\begin{itemize}
\item
 In Section 2 we define the Action and Euler-Lagrange equations for
complex non-Abelian CS gauge fields interacting with the nonrelativistic
fermions. Imposing the holomorphic gauge  we perform the Dirac
quantization.

\item
 In Section 3 we introduce     the non-unitary similarity transformation and
reduce Hamiltonian to (quasi)free form. Diagonalisation is
complete in the Abelian case.

\item
 In Section 4 we consider the system of planar electrons in external magnetic
field. As output we give the construction of the relevant wave functions for
quantum Hall fluid with Abelian as well as  non-Abelian CS gauge interactions.

\item
Section 5 containes the conclusions.

\end{itemize}

\vspace{5mm}

Together with the conventional cartezian coordinates
${\bf r}=x^k=(x,y)$
 it is convenient to use the complex notations
$$
z=x+iy,\hhh \partial =\frac{\partial}{\partial z}=\half(\partial_x-i\partial_y)
$$
$$
\bar z=x-iy,\hhh \bar \partial =\frac{\partial}{\partial
\bar z}=\half(\partial_x+i\partial_y)
 $$
for particle coordinates and corresponding Cauchy-Riemann operators.
The vector fields
${\bf A(r)}=(A_x,A_y)$
 will be  presented by their holomorphic and antiholomorphic components
$$
A({\bf r)}=A_x+iA_y,\hhh \bar A({\bf r})=A_x-iA_y
$$
The non-Abelian matrix-valued vector potential can be decomposed with
respect to a basis of the real Lie algebra of a compact gauge group $G$:
\bea
{\cal A}_\mu(x) =\sum_a A^a_\mu (x)\cdot t^a,\hh a=1,2,...,r={\rm dim}\h G
\label{eq:vpot}
 \eea
The group generators are   the anti-Hermitean,
traceless matrices  $t^a$  obeying the Lie  algebra
\bea
{t^a}^\dagger=- {t^a} ,\hhh
[t^a,t^b]=f_{abc}t^c  \label{eq:alg}
\eea
with $f^{abc}$ the totally antisymmetric, real structure constants.
In the case of abelian group (\ref{eq:vpot}) is replaced by
 ${\cal A}_\mu(x)=i A_\mu(x)$. 
 
We will abbreviate the spatial coordinates of $I^{th}$ particle ${\bf r}_I$ to
$I$, when this will not be ambiguous.

\section{Action and Quantization}

\subsection{ Complex Gauge Group and Lagrangian}

Let $G$ be a compact $r$-dimensional  Lie group.
 The group  elements are parametrized by the set of real parameters
$g=g(\omega_1,\omega_2,...,\omega_r)$. The irreducible unitary representations
of $G$ are denoted by $D_{(\sigma)}(g)\equiv D_{(\sigma)}(\omega_a)$.
Matrices $T^a_{(\sigma)}$ are the corresponding group generators.
 They  satisfy the commutation relations
\bea
\bigl[T_{(\sigma)}^a,T_{(\tau)}^b\bigr]=\delta_{\sigma \tau}f_{abc}T_{(\sigma)}^c,\hh
1\leq a,b,c \leq r
\eea
The matter fields and quantum  states in the representation
 $ D_{(\sigma)}$ are labeled by the  weight vectors ${\bf w}_{\sigma}\equiv
w^m_{\sigma} ;\h (m=1,...,R={\rm rank}\h G) $ .

Consider the  non-compact group $G_c$ ( the  complex extention of $G$),
 regarding  the group parameters $\omega_a$ as a complex quantities and with
  a group multiplication law given by the holomorphic function.
  
Recall some facts about the representations of  the complex groups.

\begin{itemize}
\item
Associated to  any irreducible
 representation $D_{(\sigma)}$ of the Lie group $G$ one
 can define its analytic and antianalytic  continuations,
  $\Delta_{(\sigma)}(g)=D_{(\sigma)}(\omega_a)$ and
  $\Delta_{(\sigma)}^\star(g) = D_{(\sigma)}(\omega^\star_a) $
  respectively.

\item
For the any two  representations $D_{(1)}$ and
$D_{(2)}$  the tensor product $\Delta_{(1,2)}(g)=
\Delta_{(1)}(g)\otimes\Delta^\star_{(2)}(g)$ is the irreducible representation
of $G_c$.
\end{itemize}

 The other irreducible representations of  interest
 are contragradient  $\tilde\Delta_{(1,2)}$,
and complex conjugate  representations
 $\Delta^\star_{(1,2)}$  and  $\tilde\Delta^\star_{(1,2)}$.
Introduce the matter fields. It is convenient to define the doublet
field
\bea
      \Psi(x)=\left( \begin{array}{c}
              \psi(x) \\
               \tilde \psi^\star(x)
              \end{array} \right), \label{eq:mdub}
 \eea
transforming under reducible representation
 ${\cal R}(g)= \Delta_{(1,2})\oplus \tilde\Delta^\star_{(1,2})$.
(The complex conjugation $^\star$ for the fermions is
defined as an  involution operation for Grassmann variables   \cite{Berezin}.)

 The corresponding contravariantly transforming fields
 $$
     \tilde \Psi(x)=\left ( \tilde  \psi(x), \psi^\star(x) \right )
$$
  are unified in the representation $\tilde{\cal R}(g)=
\tilde \Delta_{(1,2})\oplus \Delta^\star_{(1,2})$.
  It means that  there exists a nondegenerate real bilinear form
$<\tilde\Psi,\Psi>$  invariant under the group transformations.

 Gauging the rigid group $G_c$  we consider the group parameters as a
complex functions of space-time coordinates.
   The  Lie-algebra valued gauge potential
 ${\cal F}_\mu(x) \equiv F^A_\mu(x)\cdot T^A$
 transorms as follows
\bea
{\cal F}_\mu(x)\rightarrow {\cal F}^{\omega}_\mu(x)=
{\cal R}(g) {\cal F}_\mu(x){\cal R}(g)^{-1}+
\partial_\mu {\cal R}(g)\cdot {\cal R}(g)^{-1}     \label{eq:rep3}
\eea

The matrices $T^A\h (A=1,,...,r,r+1,...2r)$ are the anti-Hermitean Lie-algebra
 generators in the representation  of matter field :

\bea
T^a=\left( \begin{array}{cc}
                       T^a_{(1)}\otimes I_{(2)} & 0\\
                       0     &I_{(1)}\otimes T^a_{(2)}
                       \end{array} \right ),\hh
T^{r +a}=\left( \begin{array}{cc}
                 I_{(1)} \otimes T^a_{(2)} & 0\\
                       0     & T^a_{(1)}\otimes I_{(2)}
                       \end{array} \right )\h \label{eq:gen}
\eea
These generators are associated to the group parameters
 $\omega_a$ and  $\omega_{r+a}\equiv \omega^\star_a$ . The defining
  commutation relations are
$$
\bigl[T^A,T^B\bigr]=f_{ABC}T^C,\hh A,B=1,..,2r
$$
With the help of the gauge fields
 $ F^a_\mu (x),\h F^{r+a}_\mu (x)\equiv (F^a_\mu (x))^\star$ define the 
covariant derivatives:
\bea
 D_\mu\Psi(x)=\partial_\mu\Psi(x)-{\cal F}_\mu(x)\Psi(x) \hh
D_\mu\tilde\Psi(x)=\partial_\mu\tilde\Psi(x)+\tilde\Psi(x) {\cal F}_\mu(x) \label{eq:der}
\eea

These ingredients permit to construct the   real Lagrangian,
invariant under involution and the group of complex gauge tansformations
 $G_c$:
\bea
&&  {\cal L}=   \frac{\kappa}{2}\varepsilon ^{\mu \nu \lambda}
 [F^A_\mu(x) \partial_\nu F^A_\lambda(x) +\frac{1}{3}f_{ABC}F^A_\mu
F^B_\nu F^C_\lambda]+ \nn \\
& &\mbox{} + i<\tilde\Psi(x),D_0\Psi(x)>-
\frac{1}{2m}< D_k\tilde\Psi (x),D_k\Psi(x)> \label{eq:lhol}
 \eea
The  Euler-Lagrange equations for the matter and gauge fields are given by
the set
$$
\frac{1}{\kappa}J^A=-2i\bar \partial F^A_0-i\partial_0 F^A
-if_{ABC}F^BF^C_0
$$
\bea
\frac{1}{\kappa}\bar J^A=-2i \partial F^A_0+i\partial_0\bar F^A
+if_{ABC}\bar F^BF^C_0
           \label{eq:el}
\eea

 $$
\frac{i}{\kappa} \rho^A=\bar \partial \bar F^A-\partial F^A
-\frac{i}{4}f_{ABC}(F^B\bar F^C - \bar F^BF^C)
$$

$$
i\partial_t\psi =-\frac{1}{m}(D\bar D+\bar D D)\psi +iF_0\psi
$$
Here
\bea
D&=&\half(D_x-iD_y)=\partial+\half \bar {\cal F} \nn \\
\bar D&=&\hbox{}\half(D_x+iD_y)=\bar\partial+\half {\cal F} \nn
\eea
are covariant derivative operators.

The  gauge invariant currents
$$
J^A_0(x)\equiv \rho^A(x)=i<\tilde\Psi, T^A\Psi>
$$
$$
J^A(x)=J^A_x(x)+iJ^A_y(x)=\frac{1}{m}[<\tilde\Psi, T^A\bar D\Psi>-
<\bar D\tilde \Psi,T^A\Psi>]
$$
$$
\bar J^A=J^A_x-iJ^A_y=\frac{1}{m}[<\tilde\Psi, T^A D\Psi>-
 <D\tilde\Psi, T^A\Psi>]
$$
are covariantly conserved:
\bea
\partial_t \rho^A+\partial J^A+\bar \partial \bar J^A-   \label{eq:cont}
f_{ABC}[F^B_0\rho^C-\half (F^B\bar J^C+\bar F^B J^C)] =0
\eea

Note that the gauge coupling constant is set to be one .  Its
  actual   value can be   restored rescaling the gauge fields
      and statistical parameter
      $$  A_\mu \rightarrow gA_\mu,\hh \kappa\rightarrow \frac{\kappa}{g^2} $$

\subsection{Dirac's Quantization}

For doing canonical quantization  we will use the Dirac's  method ,
and try to adapt it for the case of complex gauge group.

To begin with, consider the classical theory
and discuss the setup of holomorphic gauge quantization.
  The canonical Hamiltonian is given by the expression
\bea
{\cal H}_c=\int d{\bf r}\Bigl[\frac{1}{m}\Bigl(<D\tilde\Psi({\bf r}) ,\bar D\Psi({\bf r}>+
<\bar D\tilde \Psi({\bf r}), D\Psi({\bf r})>\Bigr)
+F^A_0({\bf r})\phi^A({\bf r})\Bigr]    \label{eq:hc}
\eea
The system is constrained by the first class constraints
\bea
\Pi^A_0=\frac{\partial {\cal L}}{\partial \dot F^A_0}\approx 0,\nn
\eea
\bea
\phi^A=\rho^A+i\kappa\Bigl[\bar \partial \bar F^A-\partial F^A-
\frac {1}{4}f_{ABC}(F^B\bar F^C-\bar F^B F^C)\Bigr] \approx 0 \label{eq:constr}
\eea
Primary constraints reflect the absence of a momentum conjugate to $F^A_0$, and
secondary ones reproduce the generalized Gau{\ss} law.
 The canonical variables satisfy the Poisson brackets relations
\bea
\{\Psi({\bf r},t),
\tilde\Psi ({\bf r}',t)\}_{\rm PB}=
-i\delta({\bf r}-{\bf r}')
\left( \begin{array}{cc}
                 1&0 \\
                  0&1
       \end{array}  \right )             \label{eq:pb1}
 \eea
\bea
\{ F^A({\bf r} ,t),\bar F^B({\bf r} ',t)\}_{\rm PB}=
-i\frac {4}{\kappa}\delta_{AB}\delta ({\bf r-r'})     \label{eq:pb2}
 \eea

Due to the presence of quadratic terms,
the constraint equations (\ref{eq:constr}) are not easy to solve.
The obvious way out is to impose te gauge conditions, which linearize them.
This possibility is realized  in the axial type gauge {\it e.g.} $F^A_y=0$
\cite{Bak}.

As an alternative solution one can use the holomorphic gauge, with a
 gauge fixing conditions
\bea
\eta^A=F^A_0=0,\hh \chi^a=F^a=0 \hh \chi^{r+a}= \bar F^{r+a}=0
\label{eq:holg}
 \eea

In conformity with the Dirac's procedure  introduce the
 total  Hamiltonian
\bea
H_T={\cal H}_c+\int d{\bf r}\Lambda^A({\bf r})\phi^A ({\bf r})  \label{eq:ht}
\eea
where $\Lambda^A({\bf r})$
are Lagrange multipliers.
These functions must be subjected to the self-consistency conditions
\bea
\partial_t \eta^A({\bf r})=\{\eta^A({\bf r},H_T \}=0 \hh
{\rm and} \hh
\partial_t \chi^A({\bf r})=\{\chi^A({\bf r}),H_T \}=0
\eea
 and are given by
\bea
\Lambda^a({\bf r})=\frac{i}{2\kappa}\int d{\bf r}'{\cal G}
({\bf r-r'})\cdot J^{a}({\bf r}')      \label{eq:lm}
\eea
 \bea
\Lambda^{r+a}({\bf r})=-\frac{i}{2\kappa}\int d{\bf r}'\bar{\cal G}
({\bf r-r'})\cdot J^{r+a}({\bf r}')      \label{eq:lm1}
\eea
Here we formally introduced the  operator $\bar\partial^{-1}$,
which defines the Green function 
\bea
\bar \partial^{-1}J^a({\bf r})=
\int d{\bf r'}{\cal G}({\bf r-r'})J^a({\bf r'})   \label{eq:gf1}
\eea
The Green function $\bar\partial^{-1}={\cal G({\bf r})}$
 can be presented as a derivative of the holomorphic Green function
\bea
{\cal G}({\bf r})=\partial G(z)=\frac{1}{\pi z} =
\partial \frac{1}{\pi}\ln z     \label{eq:GF1}
\eea

We see, that (\ref{eq:GF1}) is ill defined multivalued function.
  In the non-relativistic case, when the particle density
   is a sum of $\delta$-functions,   using appropriate regularization
one may ignore this point and consider G(z) as a normal function,
vanishing at the origin \cite{Jackiw,Lerda}.

In the analogous way one can define the antiholomorphic  Green functions
\bea
\bar{\cal G}({\bf r})=\bar\partial G(z)=\frac{1}{\pi \bar z}   
=\frac{1}{\pi}\bar \partial\ln \bar z     \label{eq:gf2}
\eea

In the holomorphic gauge the Gau{\ss} law constraints (\ref{eq:constr})
look like
$$
\phi^a=\rho^a+i\kappa\bar \partial \bar F^a=0,\hhh
\phi^{r+a}=\rho^{r+a}-i\kappa \partial  F^{r+a}=0
$$
and can be easily   solved
\bea
\bar F^a({\bf r})=\frac{i}{\kappa}
\int d{\bf r}'{\cal G}({\bf r-r'})\cdot\rho^a({\bf r}')  \label{eq:hlpt1}
\eea
\bea
 F^{r+a}({\bf r})=-\frac{i}{\kappa}
\int d{\bf r}'\bar{\cal G}({\bf r-r'})\cdot\rho^{r+a}({\bf r}')
\label{eq:hlpt2}
\eea

In the chosen gauge  $J^a=0,\h \bar J^{r+a}=0 $.  Using the
 continuity equation (\ref{eq:cont}) one can express the Lagrange multipliers
(\ref{eq:lm}) and (\ref{eq:lm1}) as a time derivatives:
\bea
\Lambda^a({\bf r})=-\frac{i}{2\kappa}\partial_t\int d{\bf r}'G({\bf r-r'})
\rho^a({\bf r}',t)    \label{eq:lm3}
\eea
\bea
\Lambda^{r+a}({\bf r})=\frac{i}{2\kappa}\partial_t\int d{\bf r}'
\bar G({\bf r-r'})\rho^{r+a}({\bf r}',t)    \label{eq:lm4}
\eea
The last expressions may be unified with   (\ref{eq:hlpt1}) and (\ref{eq:hlpt2})
composing a 3-vectors
\bea
a^a_\mu(x)\equiv(\Lambda^a,F^a_k)=
-\frac{i}{2\kappa}\partial_\mu\int d{\bf r}'G({\bf r-r'})   \label{eq:sf1}
\rho^a({\bf r}',t)
\eea
\bea
a^{r+a}_\mu(x)\equiv(\Lambda^{r+a},F^{r+a}_k)=
\frac{i}{2\kappa}\partial_\mu\int d{\bf r}'\bar G({\bf r-r'})   \label{eq:sf2}
\rho^{r+a}({\bf r}',t)
\eea
which is the solution for the CS gauge fields in the holomorphic gauge.

\subsection {Quantum Theory}

Up to now we have  been considering the classical canonical formalism.
 The corresponding second quantized Hamiltonian operator    is given by
\bea
H_T=
\int d{\bf r}\Bigl[\frac{1}{m}\Bigl( <D\tilde\Psi({\bf r}), 
\bar D\Psi({\bf r})>+
<\bar D\tilde \Psi({\bf r}), D\Psi({\bf r})>\Bigr)+
 \Lambda^A({\bf r})\cdot \phi^A({\bf r})\Bigr] \nn
\eea

Dynamical equations are defined by the commutator
\bea
i\partial_t{\cal O}({\bf r},t)\approx\bigl [{\cal O}({\bf r},t),H_T\bigr]
\label{eq:de}
\eea
 The Heisenberg equation of motion for the matter field
 is given by
\bea
i\partial_t \Psi({\bf r},t)={\cal H}_S\Psi({\bf r},t)\equiv
-\frac{1}{m}( D\bar D
 +\bar D D ) \Psi({\bf r},t)+ia^A_0T^A\Psi({\bf r}) \label{eq:sch1}
\eea
Here the operator ${\cal H}_S$ contains the solutions  
(\ref{eq:hlpt1})-(\ref{eq:hlpt2}) for the statistical gauge fields.

It is not difficult to notice, that the many particle wave function
$$
\Phi({\bf r}_1,...,{\bf r}_N;t)=<0|\Psi({\bf r}_1,t)\cdots\Psi({\bf r}_N,t)|\Phi>
$$
satisfies the Shr$\ddot o$dinger equation
\bea
i\partial_t \Phi({\bf r}_1,...,{\bf r}_N;t) =
-\frac{1}{m}\sum_{I=1}^{N}\Bigl[ \bar D_I D_I+D_I^\star \bar D_I^\star\bigr]
\Phi({\bf r}_1,...,{\bf r}_N;t) \label{eq:sch2}
 \eea
 with the derivative operators given by
$$
D_I=\partial_I-\frac{1}{2\pi\kappa}\sum_{I\not=J}\frac{T^a_I\otimes
T^a_J}{z_I-z_J}
\hhh \bar D_I=\bar\partial_I
$$
$$
  D^\star_I=\partial_I\hhh
 \bar D^\star_I=\partial_I+\frac{1}{2\pi\kappa}\sum_{I\not=J}\frac
 {T^{r+a}_I\otimes T^{r+a}_J}{\bar z_I-\bar z_J}
$$
( the matrices $T^A_I$ act on the group variables of $I^{th}$ particle)

As a first application of the framework described above consider the case in which fermions are in the fundamental  ("chiral")  representation
$T^a_{(1)}=t^a$ and $T^a_{(2)}=0$.  The Hamiltonian takes the form
\bea
H=-\frac{1}{m}\sum_I\Bigl[\bar {\cal D}_I {\cal D}_I\cdot
\left(\begin{array}{cc}
              1&0\\
              0&0
               \end{array}\right) +{\cal D}^\star_I\bar {\cal D}^\star_I\cdot
  \left( \begin{array}{cc}
              0&0\\
              0&1
              \end{array}\right)\Bigr]
\eea
where the covariant derivative operators are given by the
KZ connections 
$$
{\cal D}_I=\partial_I-\frac{1}{2\pi\kappa}\sum_{I\not=J}\frac{t^a_I\otimes
t^a_J}{z_I-z_J}
\hhh \bar {\cal D}_I=\bar\partial_I
$$
$$
  {\cal D}^\star_I=\partial_I\hhh
 \bar {\cal D}^\star_I=\bar\partial_I+\frac{1}{2\pi\kappa}\sum_{I\not=J}\frac
 {t^a_I\otimes t^a_J}{\bar z_I-\bar z_J}
$$

Another case of interest is a "symmetric"  representation 
$T^a_{(1)}=T^a_{(2)}=t^a$.  Now
\bea
T^a=   T^{r+a}= \left( \begin{array}{cc}
              t^a&0\\
              0&t^a
              \end{array}\right)
\eea
The corresponding Hamiltonian is given by
\bea
H=-\frac{1}{m}\sum_I\Bigl[ \bar {\cal D}_I {\cal D}_I\cdot
\left( \begin{array}{cc}
              1&0\\
              0&1
               \end{array}\right) +{\cal D}^\star_I\bar {\cal D}^\star_I
  \left( \begin{array}{cc}
              1&0\\
              0&1
              \end{array}\right)\Bigr]
\eea

Remark, that in the case of "chiral" representation Hamiltonian is not
Hermitean \cite{Verlinde,Lee} and conjugation causes the interchange
between upper and down components of matter field doublet (\ref{eq:mdub}).
At the same time in the "symmetric" representation 
$H^\dagger=H$.

\section{Hamiltonian Diagonalization}

\subsection{Similarity Transformation}

The matter Hamiltonian
\bea
H_{matter}=\int d{\bf r}'\frac{1}{2m}<D_k\tilde\Psi({\bf r}),
 D_k\Psi({\bf r})>   \label{eq:ham}
\eea
containes  the  gauge connection in the form of a gradient
\bea
{\cal F}_k({\bf r})=\frac{i}{2\kappa} \partial_k\Bigl[\int d{\bf r}'
G({\bf r-r'})\rho^a({\bf r'})\cdot T^a
- \int d{\bf r}'
\bar G({\bf r-r'})\rho^{r+a}({\bf r'})\cdot T^{r+a}\Bigr] \nn
\eea

The situation is simplified when $G=U(1)$: the generators
commute and CS fields can be eliminated by the means of suitably chosen
complex gauge transformation, reducing Hamiltonian to the free (diagonal) form.

Formally the gauge fields can be removed by going to the new
field variables
$$
X (x)={\chi(x)\choose \tilde \chi^\star(x)}\hh {\rm and}\hh
\tilde X (x)=(\tilde \chi(x),\chi^\star(x))
$$
defined by
\bea
\Psi ({\bf r})= U({\bf r};\gamma)X ({\bf r})
,\hh\tilde\Psi ({\bf r})=
\tilde X ({\bf r}) U^{-1}({\bf r};\gamma)   \label{eq:si1}
\eea
where $U({\bf r};\gamma)$ is a holonomy operator
(or monodromy matrix) associated to an oriented
open path $\gamma$ in $R^2$ connecting the points ${\bf r}_0$ and
${\bf r}$:
\bea
U({\bf r};\gamma)=P \exp \Bigl( \int_{\gamma} dx^k {\cal F}_k({\bf x})\Bigr)
\label{eq:hol}
\eea
(In (\ref{eq:hol}) $P$ is path ordering operation  and ${\bf r}_0$ is
some fixed point).

Remark that, 
 due to the non-commutativity of density     operators
 $$
 [\rho^A ({\bf r}),\rho^B({\bf r}')]=f_{ABC}\rho^C({\bf r})\delta({\bf r-r'}),
 $$
path-ordering is non-trivial operation. 
Below we describe  much simpler procedure which in principle permits to
get some information on the non-Abelian wave functions.
 
Introduce operator
\bea
\Omega({\bf r})=-\frac{i}{2\kappa}\int d{\bf r'}G({\bf r-r'})
\rho^m({\bf r'})\cdot H^m
+\frac{i}{2\kappa}\int d{\bf r'}\bar G({\bf r-r'})
\rho^{r+m}({\bf r'})\cdot H^{r+m}  \nn         
\eea
where
$$
 \rho^M({\bf r})=i\tilde\Psi({\bf r})H^M \Psi({\bf r})
 $$
 are mutually commuting charge densities. The matrices
\bea
H^m= { \left( \begin{array}{cc}
              H^m_{(1)}&0\\
              0&H^m_{(2)}
              \end{array}\right)}\hh
H^{r+m}=  {\left( \begin{array}{cc}
              H^m_{(2)}&0\\
              0&H^m_{(1)}
              \end{array}\right)},\hh m=1,...,R \nn
\eea
are Cartan generators in the representation  ${\cal R}(g)$.
Consider  transformations
 \bea
\Psi ({\bf r})=  e^{\Omega ({\bf r})} X ({\bf r})
,\hh\tilde\Psi ({\bf r})=
\tilde X ({\bf r}) e^{-\Omega ({\bf r})} \label{eq:xi1}
\eea
 The action of the diagonal  Cartan generators  on the $\Psi$-fields
 \bea
H^M\Psi_{\bf w}=\Upsilon_{\bf w}^M \cdot \Psi_{\bf w}
 \label{eq:alpha}
\eea
defines  the $2R$ dimensional weight vector $\Upsilon_{\bf w}^M$
($M=1,...,R,r+1,...r+R$)
\bea
\Upsilon_{\bf w}^m= \left( \begin{array}{cc}
              w^m_1&0\\
              0& w^m_2
             \end{array}\right) ,\hh
 \Upsilon_{\bf w}^{r+m}= \left( \begin{array}{cc}
              w^m_2&0\\
              0&w^m_1
              \end{array}\right)
\eea
(Remind, that ${\bf w}_{\sigma}$'s  are the weight vectors of the
 representation $D_{(\sigma)}$).

The transformation (\ref{eq:xi1}) can be written in the  component
form
\bea \Psi_{\bf w}({\bf r})=
\sum_{\bf w'}
\Bigl(e^{\Omega({\bf r})}\Bigr)_{\bf w,\bf w'} X_{\bf w'}({\bf r})=
 e^{\Omega_{\bf w}({\bf r})}X_{\bf w}({\bf r}) \label{eq:chi1}
\eea
where the  operators
\bea
\Omega_{\bf w}({\bf r})= -\frac{i}{2\kappa}\int d{\bf r'}G({\bf r-r'})
\rho^m({\bf r'})\cdot \Upsilon_{\bf w}^m
+\frac{i}{2\kappa}\int d{\bf r'}\bar G({\bf r-r'})
 \rho^{r+m}({\bf r'})\cdot \Upsilon_{\bf w}^{r+m} \nn
 \eea
are labeled by the corresponding weight vectors.

In order to find the (anti)commutation rules obeyed by matter fields we
use the relations
\bea
[\Omega_{\bf w} (1),\Psi_{\bf w'}(2)]=
-\frac{1}{2\kappa}G(1-2)<{\bf w,w'}>
\left( \begin{array}{cc}
             1&0\\
              0&1
             \end{array}\right)  \Psi_{\bf w}(2) \label{eq:cr1}
\eea

  Suppose that the matter field $X(x)$ as well as $\Psi(x)$
corresponds to the fermions.  Straightforward calculations show, that the
requirement of the fermionic  commutation relations together with
(\ref{eq:chi1}) and (\ref{eq:cr1}) 
 leads  to the  conditions  imposed on the weight vectors of
   the representations $D_{(1)}$ and $D_{(2)}$:
\bea
e^{\frac{1}{2\kappa}<{\bf w}_1,{\bf w'}_1>}=
e^{\frac{1}{2\kappa}<{\bf w}_2,{\bf w'}_2>}=1   \label{eq:int}
\eea

Remark, that in certain circumstances it is more suitable to consider $X(x)$
as bosonic fields ({\it e.g.} in the Ginzburg-Landau description of QHE).
In these cases instead of (\ref{eq:int}) we have
to choose the weight lattice which satisfies the condition
\bea
e^{\frac{1}{2\kappa}<{\bf w}_1,{\bf w'}_1>}=
e^{\frac{1}{2\kappa}<{\bf w}_2,{\bf w'}_2>}=-1   \label{eq:int1}
\eea

 \subsection { Wave Functions}

  Let $|\Phi>$ be the eigenstate of the  Hamiltonian $H_{matter}$.
   The corresponding $N$-particle wave function  is given by
the matrix elements
$<0| \Psi_{{\bf w}(1)}({\bf r}_1)\cdots
\Psi_{{\bf w}(N)}({\bf r}_N)|\Phi>$.
After elementary manipulations one gets
\bea
& & <0| \Psi_{{\bf w}(1)}({\bf r}_1)\cdots
\Psi_{{\bf w}(N)}({\bf r}_N)|\Phi>= \nn \\
 & & \mbox{}\prod_{I<K} {\cal U}(z_I,{\bf w}(I);z_J,{\bf w}(J))
 <0|e^{ \sum\Omega_{\bf w}(I)({\bf r}_I)}
X_{{\bf w}(1)}({\bf r}_1)\cdots X_{{\bf w}(N)}({\bf r}_N)|\Phi>\nn
 \eea
Here
\bea
&&{\cal U}(z_I,{\bf w}(I);z_J,{\bf w}(J)>)=
G(z_I-z_K)
\left( \begin{array}{cc}
              w^m_1(I)\cdot w^m_1(K)& 0\\
              0& w^m_2(I)\cdot w^m_2(K)
            \end{array}\right) -  \nn  \\
&&\mbox{}- \bar G(z_I-z_K)
\left( \begin{array}{cc}
              w^m_2(I)\cdot \w^m_2(K)&0 \nn\\
              0& w^m_1(I)\cdot w^m_1(K)
             \end{array}\right)
 \eea

Following (\ref{eq:int}) the weight vectors must belong to the lattice defined
by the equations
\bea
\frac{1}{2\kappa\pi}<{\bf w}_1(I),{\bf w}_1(K)>=\pm 2p_{IK}\nn
\eea
\bea
\frac{1}{2\kappa\pi}<{\bf w}_2(I),{\bf w}_2(K)>=\pm 2q_{IK}\nn
\eea
with  $p_{IK}$ and $ q_{IK}$ integers.  (In the bosonized theory
 r.h.s. of these relations are changed by $\pm 1$ giving the odd numbers)

As we see, the wave function is factorized into  the "kinematical" prefactor
and  some dynamical part.
The typical term in the prefactor is of the form
\bea
\left(\matrix{(z_I-z_K)^{\pm 2p_{IK}}(\bar z_I-\bar z_K)^{\mp 2q_{IK}}&\cr
0&(z_I-z_K)^{\pm 2q_{IK}}(\bar z_I-\bar z_K)^{\mp 2p_{IK}}&\cr}\right)
\label{eq:pf}
\eea
and depends on the representations and  quantum numbers carried by
particles under consideration.

As a practical application of the proposed scheme one can indicate the
theory of quantum Hall effect  (see {\it e.g.}\cite{Stone}),
 where the expressions like (\ref{eq:pf}) are used as a building blocks for
the many-particle wave functions.

This and other developments will be considered in the
subsequent section.

\section{The Quantum Hall Effect and Laughlin States}

 \subsection{Abelian  Theory}

In the Abelian case which
 is obtained by the formal  substitutions  $H^A\rightarrow {\rm i},\h
 \rho^a\rightarrow -\rho=-\tilde\psi\psi,\h \alpha_A\rightarrow {\rm i} $,
the  gauge fields can be removed by the non-unitary  similarity transformations \cite{Karabali},\cite {Eliashvili}(see Appendix A):
\bea
\psi ({\bf r}) =S\chi ({\bf r}) S^{-1} =e^{-\frac{1}{2\kappa}
\int d{\bf r'}G(z-z')\tilde \chi({\bf r}')\chi ({\bf r}')}  \label{eq:chi2a}
\eea 
\bea 
\tilde\psi({\bf r})= S \tilde\chi({\bf r}) S^{-1}=
\tilde \chi({\bf r})e^{\frac{1}{2\kappa}
\int d{\bf r'}G(z-z')\tilde \chi({\bf r}')\chi ({\bf r}')} \label{eq:chi2b}
\eea 
(Here and hereafter we abandon the doublet notations). 
 The $\chi$ fields will satisfy the Fermi-Dirac  commutation
relations, {\it i.e.}
\bea \frac{1}{2\kappa}=\pm 2\pi p  \label{eq:stat} 
\eea 
with
positive integer   $p$ .

Remark, that  in the Abelian case the holomorphic $(A=0)$ and
axial $(A_y=0)$ gauges are  related by the complex gauge transformation:
\bea
A^k_{axial}=
A^k_{hol}-i\frac{1}{\kappa}\partial_k \lambda=
-\frac{1}{2\kappa}\delta_{k 1}\int d{\bf r}'
\delta(x-x')\epsilon (y-y')\varrho ({\bf r}')   \label{eq:axial}
\eea
 where
$$
\lambda ({\bf r})=\frac{1}{2\pi}\int d{\bf r}'\ln |z-z'|\varrho ({\bf r}')
$$
\bea
+\frac{i}{2\pi}\int d{\bf r}'\tan^{-1}|\frac {y-y'}{x-x'}|
\epsilon (x-x')\epsilon (y-y')\varrho ({\bf r}')    \nn
\eea

The Hamiltonian
\bea
H_{matter}=\int d{\bf r}'\frac{1}{2m}[D_k\tilde\psi D_k\psi+
 D^\star_k \psi^\star D^\star_k\tilde \psi^\star]  \nn
\eea
 can be presented in a free form:
\bea
H_{matter}=\int d{\bf r}'\frac{1}{2m}[\partial_k \tilde \chi \partial_k \chi+
{\partial_k}  \chi^\star {\partial_k}\tilde \chi^\star]
 \label{eq:ham1}
 \eea

 Remind, that  in the theory there are two pairs of
  canonically conjugate variables:   $(\chi, \tilde\chi)$  and
    $(\chi^\star, \tilde \chi^\star)$.
It can be shown (see Appendix B),
that  in the Abelian case  {\it tilde} -operation can be identified with the
Hermitean conjugation {\it i.e.}
  \bea
\tilde\chi=\chi^\dagger,\hhh
  \tilde\chi^\star=\chi^{\star\dagger}  \label{eq:cj}
\eea
(This fact will be useful in order to study the completness relations
in the corresponding Hilbert space).        
 
 For the further needs  it
 is convenient to use charge conjugate field $\chi_c\equiv \chi^\star$.
   In terms of newly introduced fields    Hamiltonian is expressed as follows
 \bea
 H_{matter}=\int d{\bf r}'\frac{1}{2m}
\partial_k\chi^\dagger({\bf r'})\partial_k\chi({\bf r'})- \int d{\bf r'}
\frac{1}{2m} \partial_k\chi_c^\dagger({\bf r'})\partial_k\chi_c({\bf r'})
+\Delta_1   \label{eq:ham2}
 \eea
{\it i.e.} it corresponds to the two types of free fermionis.  $\Delta_1$ is
a reordering
 constant, which will be specified below.
  The basic anticommutators are given by the relations
\bea
 \lbrace \chi ({\bf r}),\chi^\dagger({\bf r}')\rbrace=\delta({\bf r-r'}),\hh
\lbrace \chi_c ({\bf r}),\chi_c^\dagger({\bf r}')\rbrace=\delta({\bf r-r'}) \nn
\eea

\subsection{System in The External Magnetic Field  }

The quantum Hall effect is a condensed matter phenomenon, taking place
at low temperatures
when the planar system is exposed to the strong perpendicular  magnetic field
${\cal B}=\epsilon_{ik}\partial_i{ A}^k$.
Below we consider the standard case of external homogeneous
  magnetic field generated by the symmetric gauge potential
  ${ A}_x=-\half{\cal B} y , { A}_y=\half {\cal B} x$.
The corresponding Hamiltonian will be now
\bea
H_{matter}=\int d{\bf r}'\frac{1}{2m}
\lbrace\nabla_k\chi^\dagger({\bf r'})\nabla_k\chi({\bf r'})-
\nabla_k\chi_c^\dagger({\bf r'})\nabla_k\chi_c({\bf r'})\rbrace+\Delta_1
 \label{eq:ham3}
\eea
where the covariant derivatives are defined by
\bea
  \nabla_k \chi=(\partial_k-ie{ A}_k)\chi\hhh
  \nabla_k \chi_c=(\partial_k+ie{ A}_k)\chi_c \nn
\eea
For simplicity assume that system is spin polarized and treat
electrons as a scalar fermions.
  The fermion fields can be decomposed into the normal modes
  \bea
  \chi ({\bf r},t)=\sum_{n=0}^\infty \sum_{j=0}^{N_B-1} F_{nj}U_{nj}({\bf r})
  e^{-iE_nt}                             \label{eq:nm1}
  \eea
 \bea
  \chi_c ({\bf r},t)=\sum_{n=0}^\infty \sum_{j=0} ^{N_B-1} F^c_{nj}\bar U_{nj}
  ({\bf r})
  e^{iE_nt}                             \label{eq:nm2}
  \eea
 where $U_{nj}({\bf r})$ are the solutions of one-particle Schr$\ddot o$dinger
 equation
 $$
 -\frac{1}{2m}\nabla_k^2 U_{nj}({\bf r})=E_n U_{nj}({\bf r})
 $$
 and $E_n=\frac{|e{\cal B}|}{m} (n+\half) $ are the energy eigenvalues.
Quantity
$$
N_B=\frac{|e{\cal B}|}{2\pi}\cdot (Area)
$$
 is  a number of quantum states  per Landau level.
 $F_{nj}$ and $ F^\dagger_{nj}$ are the
 Fock space lowering and rizing Fermi operators and satisfy usual relations
$$
\{F_{nj}, F^\dagger_{ml}\}= \delta_{nm}\delta_{jl}
$$
 The same is valid for the  charge conjugate operators
$F^c_{nj}$ and $ F^{c+}_{ml}$.

The Hamilton  and angular momentum operators are given by
\bea
H_{matter}=\sum_{nj} E_n F^+_{nj}F_{nj}- \sum_{nj} E_n F^{c+}_{nj}F^c_{nj}+
\Delta_1    \label{eq:ham4}
\eea
\bea
J=\sum_{nj}[j F^+_{nj}F_{nj} -j F^{c+}_{nj}F^c_{nj}]+\Delta_2
\label{eq:mom}
\eea
In (\ref{eq:ham4})-(\ref{eq:mom}) we abbreviate
$$
\sum_{nj}\equiv \sum_{n=0}^\infty \sum_{j=0}^{N_B-1}
$$
The reordering constants
\bea
\Delta_1=\sum_{n=0}^\infty\sum_{j=0}^{N_B-1}E_n=N_B \sum_{n=0}^\infty E_n
\label{eq:const1}
\eea

\bea
\Delta_2=\sum_{n=0}^\infty\sum_{j=0}^{N_B-1}j=
\frac{N_B(N_B-1)}{2} \sum_{n=0}^\infty 1
\label{eq:const2}
\eea
are the energy and angular momentum of  state
with totally occupied one-particle excitations.

 The eigenstates of the Hamiltonian $H_m$
 are represented by the direct products
 \bea
 |N>\otimes |M_c>\sim  F^+_{n_1 j_1}\cdots F^+_{n_N j_N} |0>\otimes
   F^{c+}_{m_1l_1}\cdots F^{c+}_{m_M l_M} |0_c> \label{eq:state}
  \eea
 where the vacuum   states  are annihilated by the lowering operators
 $$
 F_{nj}|0>=F^c_{ml}|0_c>=0, \hh n,m=0,1,...; \hh j,l=0,1,..,N_B-1
 \label{eq:vac}
 $$
The state vector (\ref{eq:state}) corresponds to the energy eigenvalue
\bea
\frac{|e{\cal B}|}{m}[\sum_{i=1}^N [(n_i+\half)-
\sum_{k=1}^M (m_k+\half)]+\Delta_1  \nn
\eea
The angular momentum of this state is
\bea
J=\sum_{i=1}^N j_i-  \sum_{k=1}^M l_k +\Delta_2 \nn
\eea
Present the matter Hamiltonian as the sum
\bea
H_{matter}=H+H_c             \label{eq:ham5}
\eea
where
\bea
H= \sum _{nj}  E_n F^+_{nj}F_{nj},\hh
H_c=-\sum _{nj}E_n F^{c+}_{nj}F^c_{nj}+\Delta_1
 \label{eq:ham6}
\eea

The first term here corresponds to the particle degrees of freedom
and the second one to the holes in  in the charge conjugate sector.
 In order to justify this assertion define the states
\bea
|\Omega>=\prod_{n=0}^\infty\prod_{j=0}^{N_B-1} F^{+}_{nj}|0> \label{eq:gs1}
\eea
and
\bea
|\Omega_c>=\prod_{m=0}^\infty\prod_{l=0}^{N_B-1} F^{c+}_{ml}|0_c> \label{eq:gs2}
\eea
Together with the vacua $|0>$ and $|0_c>$ they satisfy the following relations
 \bea
 H|0>=H_c|\Omega_c>=0,\hh
 H|\Omega>=\Delta_1|\Omega>,\hh H_c|0_c>=\Delta_1 |0_c>\nn
 \eea

The elementary charged excitations with  the energy $E_n$ and angular
 momentum $j$  can be identified with the states
\bea
F^+_{nj}|0>\hh {\rm or}\hh F^c_{nj}|\Omega_c> \nn
\eea
In the same time the states
\bea
F_{nj}|\Omega>\hh {\rm or} \hh F^{c+}_{nj}|0_c>  \nn
\eea
can be interpreted as a opposite charge hole excitations with energy $-E_n$
and angular momentum $-j$.

The corresponding wavefunctions are determined by the matrix elements
\bea
<0|\chi({\bf r},t)F^+_{nj}|0>=
<\Omega_c|\chi^\dagger_c({\bf r},t)F^c_{nj}|0>=U_{nj}({\bf r})e^{-iE_nt}\nn
\eea
\bea
<0_c|\chi_c({\bf r},t)F^{c+}_{nj}|0_c>=
<\Omega|\chi^\dagger({\bf r},t)F_{nj}|\Omega>=\bar U_{nj}({\bf r})e^{iE_nt}
\nn
\eea

As a basic sets in the Hilbert space one can use the coordinate representation
vectors, which satisfy the completness relations
\bea
|0><0|+
\sum_{N\geq 1}\frac{1}{N!}\int [\prod_{1\leq I\leq N} d{\bf r}_I]
\chi^\dagger(1)\cdots\chi^\dagger (N)|0>
<0|\chi (N)\cdots\chi (1)=
1\hspace{-0.4em}1   \label{eq:compl1}
\eea
\bea
|\Omega><\Omega|+
\sum_{ N\geq 1}\frac{1}{N!}\int [\prod_{1\leq I\leq N} d{\bf r}_I]
\chi(1)\cdots\chi (N)|\Omega>
<\Omega|\chi^\dagger (N)\cdots\chi^\dagger (1)=
1\hspace{-0.4em}1    \label{eq:compl2}
\eea

The similar relations are hold in the conjugate sector. 
Notice, that validity of these completness relations is guaranteed
by the eq.(\ref{eq:cj}).

So the multi-particle
states  can be presented by the wavefunctions
\bea
<0|\chi(1)\cdots\chi(N_e)|\Phi_e>\hh {\rm or}\hh
<\Omega|\chi^\dagger(1)\cdots\chi^\dagger(N_h)|\Phi_e> \label{eq:mpwf}
\eea
and multi-hole state by the wavefunctions
\bea
<0_c|\chi_c(1)\cdots\chi_c(N_h)|\Phi_h>\hh {\rm or} \hh
<\Omega_c|\chi_c^\dagger(1)\cdots\chi_c^\dagger(N_e)|\Phi_h> \label{eq:mhwf}
\eea

 \subsection{Laughlin Wave Functions}

In the theory of QHE the distinguished role is played by the
states, where all the particles are in the lowest Landau level (LLL)  .
For the  LLL  ($n=0$) operators and wave functions we'll use simplified
notations
$$
 F_{0j}\equiv f_j,\hhh U_{0j}({\bf r})\equiv u_j({\bf r})
$$
Decompose
$$
\chi ({\bf r})  =\chi_0({\bf r}) +\chi'({\bf r})
$$
where
$$
\chi_0({\bf r})=\sum_{j=0}^{N_B-1} f_ju_j({\bf r})
$$
is the lowest level field operator and LLL states are builded up by the
application of lowering and rizing operators satisfying the oscilltor algebra:
 $$
 \{f_j, f^\dagger_l\}=\delta_{jl}.
$$
 Totally filled LLL state is presented by the vector
$$
|\omega>=\prod_{0\leq j\leq N_B-1}f^\dagger_j|0>
$$
The analogous state in conjugate sector will be given by
$$
|\omega_{c}>=\prod_{0\leq j\leq N_B-1} f_j^{c\dagger}|0_c>
$$

 Instead of identity resolution (\ref{eq:compl2})  for the LLL states
one can use the  LLL projection operator
\bea
\Pi &=&|\omega><\omega,|+ \label{eq:proj1}  \\
& &\mbox{}+\sum_{1\leq N\leq N_B}\frac{1}{N!}\int [\prod_{1\leq I\leq N}
 d{\bf r}_I] \chi(1)\cdots\chi (N)|\omega> <\omega|\chi^\dagger
(N)\cdots\chi^\dagger (1)      \nn
 \eea
 and its conjugate partner
\bea
\Pi_c &=&|\omega_c,N_B><\omega_c,N_B|+ \label{eq:proj2} \\
& &\mbox{}+\sum_{1\leq N\leq N_B}\frac{1}{N!}\int [\prod_{1\leq I\leq N}
 d{\bf r}_I] \chi_c(1)\cdots\chi_c (N)|\omega_c> <\omega_c|\chi_c^\dagger
(N)\cdots\chi_c^\dagger (1)  \nn
\eea

The eigenstates of  Hamiltonian (\ref{eq:ham4}) are expressed in terms
of $\chi$  quanta excitations. At the same time the physical
observables and wave functions must be expressed in terms of the
 fields $\psi$. As we  have already noted, these operators
  are related by the similarity transformations 
({\ref {eq:chi2a})-(\ref{eq:chi2b}).
In terms of $\psi$ fields the completness relations and projection
opertors are given by
\bea
& & S|0><0|S^{-1}+  \label{eq:compl3a} \\
&  &\mbox{}+\sum_{N\geq 1}\frac{1}{N!}\int [\prod_{1\leq i\leq N} d{\bf r}_i]
\tilde\psi(1)\cdots\tilde\psi (N)S|0>
<0|S^{-1}\psi (N)\cdots\psi (1)=1\hspace{-0.4em}1 \nn
\eea
\bea
&  &S |0_c><0_c| S^{-1}+  \label{eq:compl4a} \\
&  &\mbox{}+\sum_{N\geq 1}\frac{1}{N!}\int [\prod_{1\leq i\leq N} d{\bf r}_i]
 \tilde\psi^\star(1)\cdots\tilde\psi^\star (N) S|0_c>
<0_c| S^{-1}\psi^\star (N)\cdots\tilde\psi^\star (1)=
1\hspace{-0.4em}1_c     \nn
\eea
 \bea
\Pi&=&S|\omega><\omega|S^{-1}+ \label{eq:proj3a} \\
& &\mbox{}+\sum_{1\leq N\leq N_B}\frac{1}{N!}\int [\prod_{1\leq I\leq N} d{\bf
r}_I] \psi(1)\cdots\psi (N)S|\omega> <\omega|S^{-1}\tilde\psi
(N)\cdots\tilde\psi (1) \nn
\eea
\bea
\Pi_c&=& S|\omega_c><\omega_c| S^{-1}+ \label{eq:proj4a}\\
& & \mbox{}+\sum_{1\leq N\leq N_B}\frac{1}{N!}\int [\prod_{1\leq I\leq N} d{\bf
r}_I] \psi^\star(1)\cdots\psi^\star (N) S|\omega_c> <\omega_c|
S^{-1}\tilde\psi^\star (N)\cdots\tilde\psi^\star (1) \nn
\eea

 Equations (\ref{eq:compl3a}-\ref{eq:proj4a}) together with
the properties of similarity transformation can be used in order
to make a reasonable choice of the Hilbert space basis.  
 Below we list these sets indicating 
 corresponding coordinate representation bra-vectors.

1. Vacua are invariant under similarity transformation
\bea
 S|0>=|0>,\h <0|S^{-1}=<0|\hh \rightarrow\hh <0|\psi(1)\cdots\psi(N)  \label{eq:bas1}
\eea
\bea
  S|0_c>=|0_c>,\hh <0_c| S^{-1}=<0_c|\hh \rightarrow \hh 
 <0_c|\psi^\star(1)\cdots\psi^\star(N)
    \label{eq:bas2}
\eea

2. S operator does not cause the Landau level mixing
\bea
S\Pi S^{-1}=\Pi \hh \rightarrow \hh 
<\omega|S^{-1}\tilde\psi (1)
\cdots\tilde\psi (N)  
    \label{eq:bas3}
\eea
\bea
S\Pi_c S^{-1}=\Pi_c \hh \rightarrow \hh
<\omega_c |S^{-1}\tilde\psi^\star (1)
\cdots\tilde\psi^\star(N) \label{eq:bas4}   
\eea

The LLL projected Hamiltonian is
\bea
H_0={\cal H}+{\cal H}_c \label{eq:ham7}
\eea
where
\bea
{\cal H}=E_0\sum_{j=0}^{N_B-1}f^+_j f_j \nn
\eea
and
\bea
{\cal H}_c=-E_0\sum_{j=0}^{N_B-1}f^{c+}_j f^c_j + N_B E_0 \nn
\eea
The corresponding  angular momentum operator is given by
  \bea
J_0=\sum_{j=0}^{N_B-1} j[f^+_j f_j -f^{c+}_j f^c_j]+\frac{N_B(N_B-1)}{2}\nn
\eea
Consider the state
\bea
|\Phi;N_e>=\prod_{j=0}^{N_e-1}f^+_j|0>        \label{eq:LL}
\eea

 This state   corresponds to the system of $N_e$ electrons in LLL
 with the energy  $N_e  E_0$ and minimal total angular momentum
$  J=\half N_e(N_e-1) $

The supplementary state in the conjugate sector
\bea
|\Phi_c;N_h>=\prod_{j=0}^{N_h-1}f^{c+}_j|0_c>        \label{eq:LLc}
\eea
describes the system of $N_h=N_B-N_e$ holes with the same total energy end
angular momentum
$$
 J_c= \half N_B(N_B-1)- \half N_h(N_h-1)
$$
Consequently
\bea
|L;N_e>=|\Phi;N_e>\otimes |\omega_c>    \label{eq:LS}
\eea
and
\bea
|G;N_e>=|0>\otimes |\Phi_c;N_h>    \label{eq:GS}
\eea
 are degenerate eigenstates of $H_0$ with energy $N_e E_0$.

Now it is easy to show, that (\ref{eq:LS})  describes the 
 Laughlin state \cite{Laughlin}  with filling fraction
$$
\nu= \frac{1}{2p+1}
$$
 The corresponding wave function  is obtained
 applying the projection operator $\Pi$:
\begin{eqnarray}
1\hspace{-0.4em}1\otimes \Pi |L;N_e>&\rightarrow& \Psi_\nu^e(1,,,,,N_e)
=<0|\psi(1)\cdots\psi(N_e)|\Phi,N_e> \nn \\
&=& \prod_{K<  L}(z_K-z_L)^{2p}
\langle 0|\chi (1)\cdot \cdot
\cdot \chi (N_e)|\Phi;N_e \rangle
\end{eqnarray} \label{eq:laugh2}
 The last factor
$$
\langle 0|\chi (1)\cdots \chi (N_e) f^+_{0}\cdots  f^+_{N_e-1}
|0\rangle = \prod_{1\leq K<L\leq N_e}(z_K-z_L)e^{-\frac{e{\cal B}}{4}
\sum_{I=1}^{N_e}|z_I|^{2}}
$$
 is the Slater determinant of one-particle LLL states.

Another state of interest is the Girvin state (\ref{eq:GS}). The corresponding
wave function is extracted acting by the projection operator $\Pi_c$:
  \bea
1\hspace{-0.6em}1\otimes \Pi_c |G;N_e>\rightarrow  \Psi_{\nu_c}^e(1,...,N_e)
=\langle \omega_c|S^{-1}\tilde\psi^\star(1)\cdots
\tilde\psi^\star(N_e) |\Phi_c\rangle =  \label{eq:givr1}
\eea
$$
 \int\cdots\int \prod_{K=N_e}^{N_B} [d{\bf r}_K]
\langle \omega_c | S^{-1}\tilde \psi^\star(1) \cdots
\tilde\psi^\star(N_B)|0_c\rangle \times
$$
$$
\langle 0_c|\psi^\star(N_B)\cdots \psi^\star(N_e+1)|\Phi_c>=
$$
 $$
=\int\cdots\int \prod_{K=1}^{N_e} [d{\bf r}_K] \overline
{\Psi^e_0(1,\cdots,N_B)}\times \Psi^e_p(1,\cdots,N_e)
 $$
where the relation
$
<\omega_c| S^{-1}= <\omega_c|
$
is assumed to be valid.
This wave function  describes the state with
filling fraction $\nu_c=1-\nu=2p/2p+1$  \cite{Girvin}
($\Psi^e_0$ corresponds to the totally
filled lowest level). 

Another representation of the same state will be given by the matrix element
$$
\langle \omega|S^{-1}\tilde\psi(N_e+N_h)\cdots
\tilde\psi (N_e+1) |\Phi \rangle =
 $$
$$ \int\cdots\int \prod_{K=1}^{N_h} [d{\bf r}_K]
\langle \omega|S^{-1}\tilde \psi^(N_e+N_h) \cdots
\tilde\psi(N_e+1)
\tilde\psi^(N_e)\cdots \tilde\psi^(1)|0\rangle
\times
$$
 $$
 \langle 0|\psi (1)\cdots\psi (N_e) |\omega\rangle =
 $$
 \bea
\int\cdots\int \prod_{K=1}^{N_e} [d{\bf r}_K] \overline
{\Psi^e_0(1,\cdots,N_B)}\times \Psi^e_p(1,\cdots,N_e)  \label{eq:givr2}
 \eea

The holomorphic factor $\prod (z_I-z_J)^{2p}$ is usually associated
to the $2p$ magnetic flux quanta attached to electrons forming
what is called Jain's composite particles \cite{Jain}. In the 
present discussion it is a matrix element of complex gauge transformation
relating two different, non-unitary equivalent basis.
  
In the same fashion one can consider the wave functions for
the  noncompressible states   of fractionally charged  quasiparticles.

 \subsection {Non-Abelian Wave Functions}

  Although in the  non-Abelian case we do not know the exact wave function,
 one can neverthless get some sort
of kinematical information, contained in the form of similarity
transformations (\ref{eq:chi1}).  
 In order to find the $N$-particle wave function
we need some basis vectors, {\it e.g.}
\bea
& & <0| \psi_{\bf w(1)}(1)\cdots\psi_{\bf w(N)}(N)=  \label{eq:nawf}\\
& & \mbox{}<0|\exp [\sum\Omega_{{\bf w}(I)}(I)]\exp 
{[\frac{1}{2\kappa}\sum_{I>K} G(z_I-z_K)<{{\bf w}(I)}, {{\bf w}(K)}>]}\nn\\
&&\mbox{}<0|\chi_{{\bf w}(1)}(1)\cdots\chi_{{\bf w}(N)}(N)\nn \\
& &\mbox{}=\prod_{I<K}(z_I-z_K)^{-\frac{1}{2\pi\kappa}
<{\bf w}(I), {\bf w}(K)>} 
 <0|\chi_{{\bf w}(1)}(1)\cdots\
\chi_{{\bf w}(N)}(N) \nn
\eea
 (deriving (\ref{eq:nawf}) we've used the fact, that vacuum is annihilated
by operators $\Omega_{{\bf w}(I)}(I)$).

Apply this formula to the case of $SU(2)$ non-Abelian theory.
For the $\psi  $'s  in the fundamental representation the weight vectors
$$
w^\alpha =\pm i 
$$
correspond to the isospin  up $\uparrow$ an down $\downarrow$ components.
The corresponding  basis vector is given by the expression
\bea
\prod (z_{I\uparrow}-z_{K\uparrow})^{2p}
\prod (z_{R\downarrow}-z_{S\downarrow})^{2p}
\prod (z_{I\uparrow}-z_{R\downarrow})^{-2p}
<0|\chi_{{\bf w}(1)}(1)\cdots \chi_{{\bf w}(N)}(N)   \label {eq:hwf}
\eea

So we see that 
the wave function of any Hamiltonian eigenstate in this basis
containes an holomorphic prefactor
indicating the attraction between different isospins and repulsion between the
same ones.
It was conjectured that this type of wave functions may
be related to the multilayered QHE states \cite{Halperin,Ezawa}

\section{Conclusions}

In summary, we have seen that Chern-Simons theory with
complex gauge group and doubled number of matter 
as well as gauge degrees of freedom
have natural application to the description of quantum Hall effect.
The additional matter fields, introduced in order to provide
unitarity requirement are associated to quantum states with
the observed values of filling fractions. 
The scheme incorporates also the Jains picture \cite{Jain} of composite
fermions.
 
 In the case of non-Abelian interactions 
one can extract from the complete wave function a certain
"kinematical" prefactor, having specific holomorphic form which is
determined be the weight lattice
and possibly is related to the multilayered planar system.
The bosonized version of similarity transformation can be used
to construct the order parameter and to develop 
the GL description of quantum fluids.

\section{Acknowlegments}

I would like to acknowlege the hospitality of the 
Institut f\"ur Theoretishe Physik der Universit\"at  Z\"urich,
where this work was completed. This work was supported by the
Swiss National Science Foundation.

\section{Appendix}

\appendix

\section{Similarity Transformation}

 The generator of similarity transformation must satisfy the commutation
relation
$$
[G_p,\chi ({\bf r})]=2p\pi\int d{\bf r'} G (z-z')\varrho ({\bf r}')\cdot
 \chi ({\bf r})
$$
Present $G_p$  as a bilinear functional of a density operator:

$$
G_p=\int d{\bf r}' \int d{\bf r}''\varrho ({\bf r}')\Lambda_p
 ({\bf r}',{\bf r}'')\varrho ({\bf r}'')+
\int dr L_p({\bf r})\varrho ({\bf r})
$$
Using the commutator
$$
[\varrho ({\bf r}'),\chi ({\bf r})]=-
\delta ({\bf r}-{\bf r}')\chi ({\bf r}),
$$
we get
\bea
\int dr'[\Lambda_p ({\bf r},{\bf r}')+\Lambda_p ({\bf r}',{\bf r})]
\varrho ({\bf r}')=
-2p\pi\int d{\bf r}' G (z-z')\varrho ({\bf r}'),\nn
\eea
and
$$
L_p({\bf r})=-\Lambda_p ({\bf r},{\bf r}).
$$
Consequently
$$
\Lambda_p ({\bf r},{\bf r}')=-p\pi G (z-z')+\frac {i\pi}{2}p,
$$
and
$$
L_p({\bf r})=-\lim_{{\bf r}'\rightarrow {\bf r}}
\Lambda_p ({\bf r},{\bf r}')=i\frac{\pi}{2}p
$$
In the last expression we have used the regularized Green function,
satisfying condition $G(0)=0$.
As an example of such a regularization one can try to use a function
$$
G(z)=\lim_{\epsilon\rightarrow 0}G_{\epsilon}(z),\hh
 G_\epsilon=\frac{1}{\pi}\ln z\cdot e^{-\epsilon/|z|^2}
$$

\section{Toy Model}

Consider a toy model describing the couple of fermion oscillators.
The basic anticommutators are
\bea
\{ f,f^\dagger\}=1,\hhh \{ f_c,f_c^\dagger \}=1  \nn
\eea
The Hamiltonian is
\bea
\hat H=f^\dagger f-f_c^\dagger f_c              \nn
\eea
The explicit realization of basic operators and Hilbert space can be given
in terms of Grassmann variables $\xi$ and $\xi^\ast$:
\bea
f=\frac{\partial}{\partial \xi}, \hh f^\dagger =\xi\hhh
f_c=\frac{\partial}{\partial \xi^\ast}, \hh f_c^\dagger =\xi^\ast    \nn
\eea
Introduce the dual vector
\bea
\Psi^{\#} =-\bar\psi_{00}\xi\xi^\ast+\bar\psi_{01}\xi^\ast-\bar\psi_{10}\xi-
\bar\psi_{11}                                                              \nn
\eea
The scalar product is defined by Berezin the integral over
the Grassmann numbers
\bea
& &(\Phi,\Psi)=\int {\rm d}\xi {\rm d} \xi^\ast \Phi^{\#}\cdot \Psi=\nn\\
&& \mbox{}=\bar\phi_{00}\psi_{00} +\bar\phi_{10}\psi_{10} +
\bar\phi_{01}\psi_{01} +\bar\phi_{11}\psi_{11}                     \nn
\eea
We see, that the  pairs of Hermitian conjugate operators are given by
\bea
(\xi, \h \xi^\dagger=\frac{\partial}{\partial \xi}) \hh{\rm and} \hh
(\xi^\star,\h \xi^{\ast\dagger}=\frac{\partial}{\partial \xi^\ast})                 \nn
\eea

 So our Hamiltonian is Hermitian and invariant under involution operation.

\end{document}